\documentclass{article}
\usepackage[preprint]{neurips_2021}

\usepackage[hyphens]{url}
\usepackage[colorlinks=false,hidelinks]{hyperref}
\usepackage{graphicx}
\usepackage{tabularx}
\usepackage{booktabs}
\usepackage{amssymb}
\usepackage{amsmath}
\usepackage{amsthm}
\usepackage{mathtools}
\usepackage{listings}
\usepackage{multirow}

\setcitestyle{numbers,square}

\title{JSON Stats Analyzer}

\author{
  Juan Cruz~Viotti\thanks{\url{https://www.jviotti.com}} \\
  Department of Computer Science \\
  University of Oxford \\
  Oxford, GB OX1 3QD \\
  \texttt{jv@jviotti.com} \\
  \and
  Mital~Kinderkhedia \\
  Department of Computer Science \\
  University of Oxford \\
  Oxford, GB OX1 3QD \\
  \texttt{mital.kinderkhedia@cs.ox.ac.uk} \\
}

\begin{document}
\maketitle

\begin{abstract}

In this paper, we present the JSON Stats Analyzer, a free-to-use open-source
  web-based JavaScript tool and module that provides JSON document analysis. We
  explain how the JSON Stats Analyzer works, its usage alongside the
  demonstration of eleven JSON documents from Tier 1, Tier 2 and Tier 3 from
  our proposed taxonomy that categorizes JSON documents according to their
  size, content, redundancy and nesting characteristics. For each JSON
  document, we provide its definition, characteristics and the document
  structure, alongside a visual representation of the JSON document structure
  and its summary statistics.

\end{abstract}

\section{Motivation}

To the best of our knowledge, there are gaps in the current literature that
prevents a sound understanding of JSON \cite{ECMA-404} documents at a higher
level that provides us with an insight of the instances themselves.  As a
consequence, research experiments that involve JSON documents do not follow a
methodical approach for solving the input selection problem.  To solve this
problem, \cite{viotti2022benchmark} (Table 2) introduces a formal tiered
taxonomy consisting of 36 categories that form a common basis to class JSON
documents based on their size, type of content, characteristics of their
structure and redundancy criteria.

Given the universality of the JSON \cite{ECMA-404} data interchange format as
the lingua-franca of the web \cite{10.1145/3355369.3355594}, a growing amount
of research is conducted taking JSON documents as input. For any experiment
following a thorough scientific approach, the quality and representativity of
the input data is a significant factor that determines the quality of the
experiment results.

Software systems make use of JSON to model diverse and domain-specific data
structures. Each of these data structures have characteristics that distinguish
them from other data structures. For example, a data structure that models a
person is fundamentally different from a data structure that models sensor
data.  The difference lies in the characteristics which use diverse data types
such as text, numeric and boolean. Therefore, two instances of the same data
structure may inherit the same or similar characteristics despite having
different values.

Intuitively and on observation, we know that these characteristics exist, but
in the existing literature we lack a common terminology to describe them in an
unambiguous way. In an attempt to solve this ambiguity problem, our taxonomy
\cite{viotti2022benchmark} (Table 2) presents a formal vocabulary to describe,
reason, discuss and reason about JSON documents in a high-level manner.

\section{Taxonomy Definition}

Our proposed taxonomy \cite{viotti2022benchmark} aims to classify JSON
documents into a limited and useful set of categories that is easy to reason
about rather than exhaustively considering every possible aspect of a data
structure. This taxonomy categorizes JSON documents according to their size,
content, redundancy and nesting characteristics.

\subsection{Naming Conventions}

We present the naming conventions of the taxonomy in
\autoref{table:naming-conventions}.

\begin{table*}[h!]

\caption{The letter \textit{T} stands for \textit{Tier 1}, the letter
\textit{S} stands for \textit{Tier 2} and the letter \textit{L} stands for
\textit{Tier 3}.}

\label{table:naming-conventions}
\begin{tabularx}{\linewidth}{Xllll}
\toprule
\textbf{Tier} & \textbf{Content Type} & \textbf{Redundancy} & \textbf{Structure} & \textbf{Acronym} \\
\midrule

\multicolumn{5}{l}{\textbf{Size Minified $<$ 100 bytes}} \\ \hline

Tier 1 & Numeric & Redundant & Flat           & \texttt{TNRF} \\ \hline
Tier 1 & Numeric & Redundant & Nested         & \texttt{TNRN} \\ \hline
Tier 1 & Numeric & Non-Redundant & Flat       & \texttt{TNNF} \\ \hline
Tier 1 & Numeric & Non-Redundant & Nested     & \texttt{TNNN} \\ \hline
Tier 1 & Textual & Redundant & Flat           & \texttt{TTRF} \\ \hline
Tier 1 & Textual & Redundant & Nested         & \texttt{TTRN} \\ \hline
Tier 1 & Textual & Non-Redundant & Flat       & \texttt{TTNF} \\ \hline
Tier 1 & Textual & Non-Redundant & Nested     & \texttt{TTNN} \\ \hline
Tier 1 & Boolean & Redundant & Flat           & \texttt{TBRF} \\ \hline
Tier 1 & Boolean & Redundant & Nested         & \texttt{TBRN} \\ \hline
Tier 1 & Boolean & Non-Redundant & Flat       & \texttt{TBNF} \\ \hline
Tier 1 & Boolean & Non-Redundant & Nested     & \texttt{TBNN} \\ \hline

\multicolumn{5}{l}{\textbf{Size Minified $\geq$ 100 $<$ 1000 bytes}} \\ \hline

Tier 2 & Numeric & Redundant & Flat           & \texttt{SNRF} \\ \hline
Tier 2 & Numeric & Redundant & Nested         & \texttt{SNRN} \\ \hline
Tier 2 & Numeric & Non-Redundant & Flat       & \texttt{SNNF} \\ \hline
Tier 2 & Numeric & Non-Redundant & Nested     & \texttt{SNNN} \\ \hline
Tier 2 & Textual & Redundant & Flat           & \texttt{STRF} \\ \hline
Tier 2 & Textual & Redundant & Nested         & \texttt{STRN} \\ \hline
Tier 2 & Textual & Non-Redundant & Flat       & \texttt{STNF} \\ \hline
Tier 2 & Textual & Non-Redundant & Nested     & \texttt{STNN} \\ \hline
Tier 2 & Boolean & Redundant & Flat           & \texttt{SBRF} \\ \hline
Tier 2 & Boolean & Redundant & Nested         & \texttt{SBRN} \\ \hline
Tier 2 & Boolean & Non-Redundant & Flat       & \texttt{SBNF} \\ \hline
Tier 2 & Boolean & Non-Redundant & Nested     & \texttt{SBNN} \\ \hline

\multicolumn{5}{l}{\textbf{Size Minified $\geq$ 1000 bytes}} \\ \hline

Tier 3 & Numeric & Redundant & Flat           & \texttt{LNRF} \\ \hline
Tier 3 & Numeric & Redundant & Nested         & \texttt{LNRN} \\ \hline
Tier 3 & Numeric & Non-Redundant & Flat       & \texttt{LNNF} \\ \hline
Tier 3 & Numeric & Non-Redundant & Nested     & \texttt{LNNN} \\ \hline
Tier 3 & Textual & Redundant & Flat           & \texttt{LTRF} \\ \hline
Tier 3 & Textual & Redundant & Nested         & \texttt{LTRN} \\ \hline
Tier 3 & Textual & Non-Redundant & Flat       & \texttt{LTNF} \\ \hline
Tier 3 & Textual & Non-Redundant & Nested     & \texttt{LTNN} \\ \hline
Tier 3 & Boolean & Redundant & Flat           & \texttt{LBRF} \\ \hline
Tier 3 & Boolean & Redundant & Nested         & \texttt{LBRN} \\ \hline
Tier 3 & Boolean & Non-Redundant & Flat       & \texttt{LBNF} \\ \hline
Tier 3 & Boolean & Non-Redundant & Nested     & \texttt{LBNN} \\

  \bottomrule
\end{tabularx}
\end{table*}

\begin{figure*}[ht!]
\frame{\includegraphics[width=\linewidth]{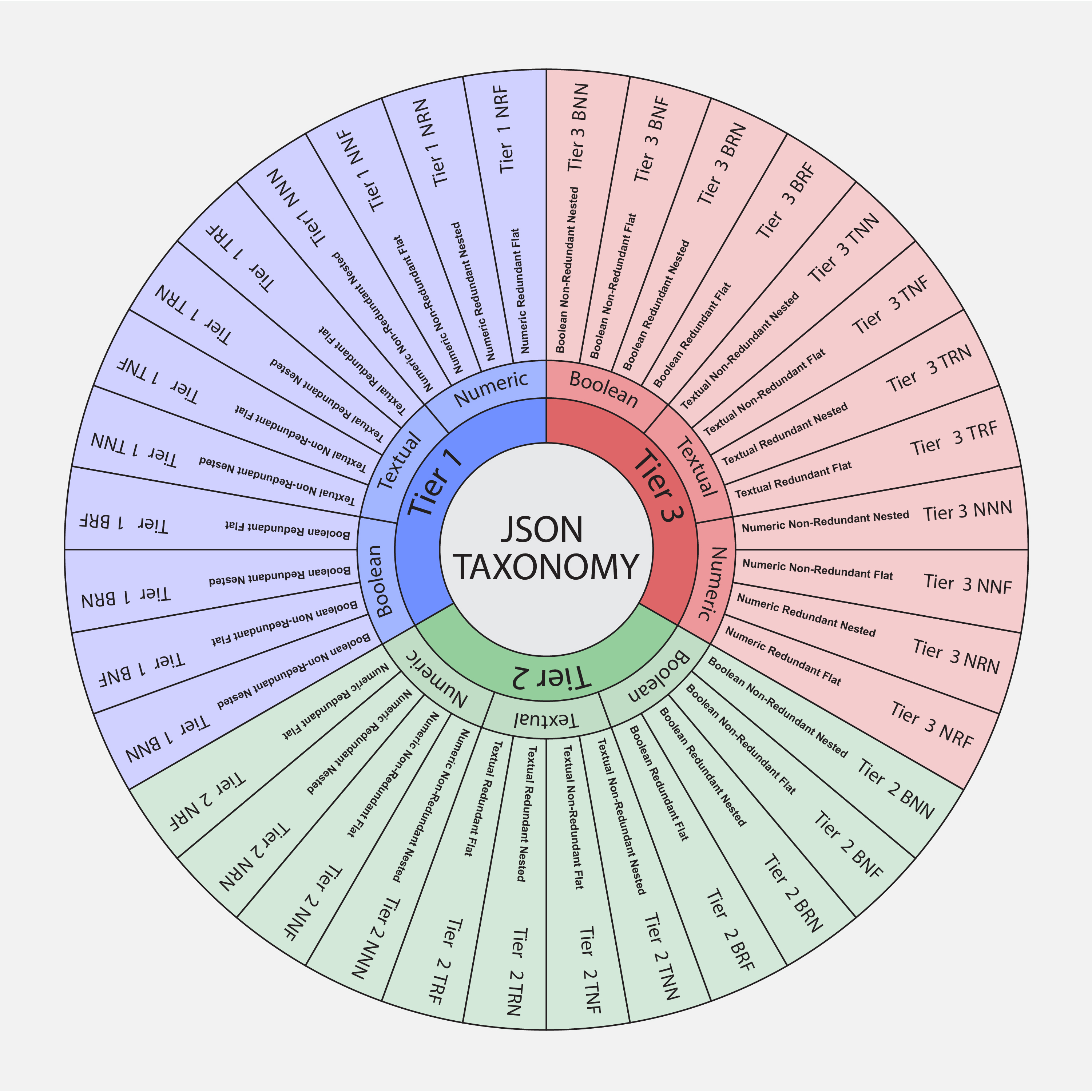}}
\caption{\cite{viotti2022benchmark} introduces a formal taxonomy to class JSON
documents that consists of 36 categories.} \label{fig:json-taxonomy}
\end{figure*}

\subsection{Size}

\begin{itemize}

  \item \textbf{Tier 1 Minified $<$ 100 bytes.} A JSON document is in this
    category if its UTF-8 \cite{UnicodeStandard} minified form occupies less
    than 100 bytes.

  \item \textbf{Tier 2 Minified $\geq$ 100 $<$ 1000 bytes.} A JSON document is
    in this category if its UTF-8 \cite{UnicodeStandard} minified form occupies
    100 bytes or more, but less than 1000 bytes.

  \item \textbf{Tier 3 Minified $\geq$ 1000 bytes.} A JSON document is in this
    category if its UTF-8 \cite{UnicodeStandard} minified form occupies 1000
    bytes or more.

\end{itemize}

\subsection{Content Type}

\begin{itemize}

  \item \textbf{Textual.} A JSON document is in this category if it has at
    least one string value and its number of string values multiplied by the
    cummulative byte-size occupied by its string values is greater than or
    equal to the boolean and numeric counterparts.

  \item \textbf{Numeric.} A JSON document is in this category if it has at
    least one number value and its number of number values multiplied by the
    cummulative byte-size occupied by its number values is greater than or
    equal to the textual and boolean counterparts.

  \item \textbf{Boolean.} A JSON document is in this category if it has at
    least one boolean or null value and its number of boolean and null values
    multiplied by the cummulative byte-size occupied by its boolean and null
    values is greater than or equal to the textual and numeric counterparts.

  \item \textbf{Structural.} A JSON document is in this category if it does not
    include any string, boolean, null or number values.

\end{itemize}

A JSON document can be categorized as textual, numeric and boolean at the same
time.

\subsection{Redundancy}

\begin{itemize}

  \item \textbf{Non-redundant.} A JSON document is in this category if less
    than 25

  \item \textbf{Redundant.} A JSON document is in this category if at least 25%
    percent of its scalar and composite values are redundant.

\end{itemize}

\subsection{Nesting}

\begin{itemize}

  \item \textbf{Flat.} A JSON document is in this category if the height of the
    document multiplied by the non-root level with the largest byte-size when
    taking textual, numeric and boolean values into account is less than 10. If
    two levels have the byte size, the highest level is taken into account.

  \item \textbf{Nested.} A JSON document is in this category if it is
    considered structural and its height is greater than or equal to 5, or if
    the height of the document multiplied by the non-root level with the
    largest byte-size when taking textual, numeric and boolean values into
    account is greater than or equal to 10. If two levels have the byte size,
    the highest level is taken into account.

\end{itemize}

\section{JSON Stats Analyzer}

The JSON Stats Analyzer is a free-to-use open-source web application that
implements the taxonomy introduced in \cite{viotti2022benchmark} and provides
summary statistics. Using JSON Stats Analyzer, a user can determine the summary
statistics corresponding to the taxonomy for any given JSON document.  To
complement the web application, the project also distributes a Node.js
\footnote{\url{https://nodejs.org}} command-line interface program and a
JavaScript library for programmatic use.

The JSON Stats Analyzer is developed using the JavaScript \cite{ECMA-262}
programming language, the CodeMirror \footnote{\url{https://codemirror.net}}
version 5.65.0 open-source embeddable web editor (MIT License), the open-source
Chart.js \footnote{\url{https://www.chartjs.org}} version 3.7.0 charting
JavaScript library (MIT License), and the UI Kit
\footnote{\url{https://getuikit.com}} version 3.10.0 CSS open-source web
component framework (MIT License). The web application is deployed to the
GitHub Pages \footnote{\url{https://pages.github.com}} free static-hosting
service.

To make use of the JSON Stats Analyzer web application, the user opens
\url{https://sourcemeta.github.io/json-taxonomy/} on a web browser, writes or
copy-pastes a JSON document of choice on the text editor at the left-hand side
of the screen, and clicks the \textit{Analyze JSON} button at the top right
corner of the screen. The analysis results are presented at the right-hand side
of the screen. If the syntax of the JSON document is invalid, an error is
displayed at the right-hand side of the screen.

The analysis results are presented in the following order:

\begin{itemize}

\item The classification of the input JSON document according to the taxonomy
  introduced in \cite{viotti2022benchmark}.

\item A set of pie charts representing the distribution of content type when
  taking into consideration the number of values of the input JSON document and
    the byte-size of the input JSON document in UTF-8 \cite{UnicodeStandard}
    minified form.

\item A high-level summary of the input JSON document in terms of its byte-size
  in UTF-8 \cite{UnicodeStandard} minified form, its number of values, its
    height and its number of duplicated values.

\item A breakdown of the input JSON document values aggregated by content and
  divided in terms of the number of values in the document, their byte-size and
    redundancy level.

\item A breakdown of the input JSON document values aggregated by their nesting
  level divided in terms of the number of values in the document and their
    byte-size.

\end{itemize}

\section{Usage}

\subsection{JavaScript}

\begin{figure*}[ht!]
\frame{\includegraphics[width=\linewidth]{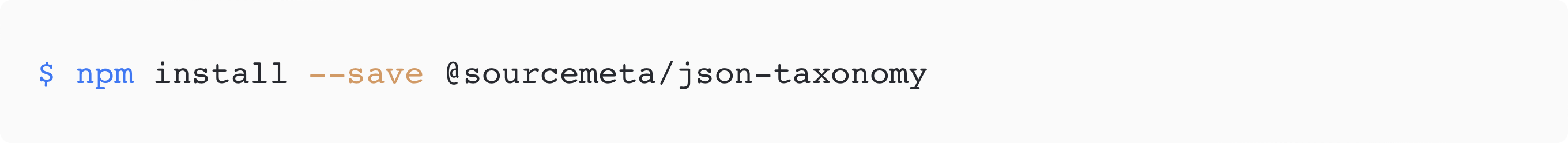}}
\caption{The JavaScript module that implements the JSON Taxonomy is installed through the npm package
manager.} \label{fig:npm-install-save} \end{figure*}

We publish an npm \footnote{\url{https://www.npmjs.com}} package which can be
installed as shown in \autoref{fig:npm-install-save}.

\begin{figure*}[ht!]
  \frame{\includegraphics[width=\linewidth]{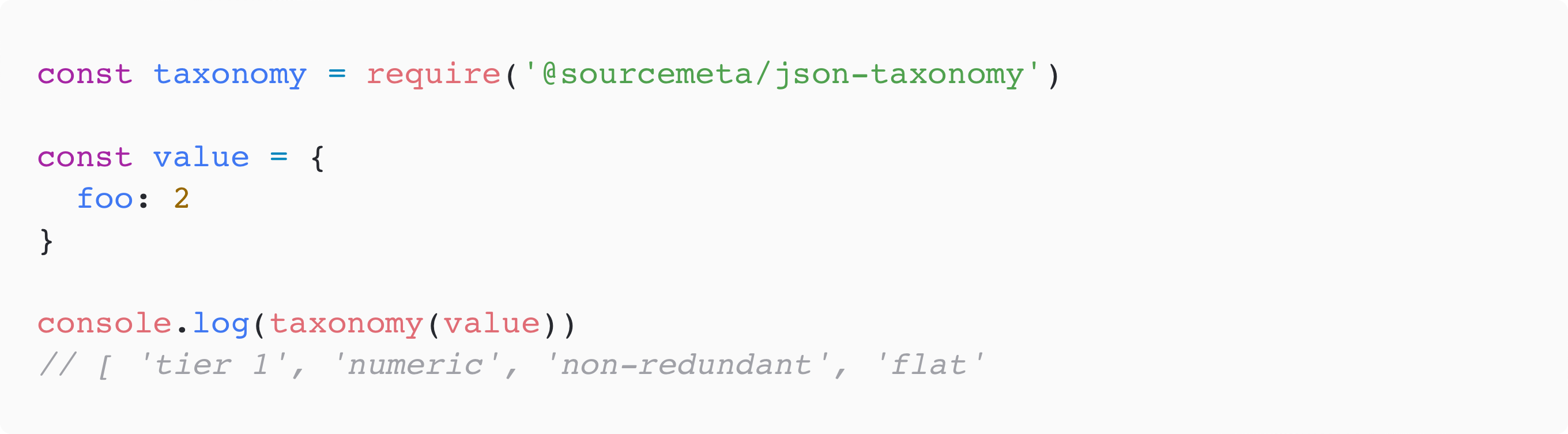}}
  \caption{The JavaScript module exposes a single function that computes the
  JSON Taxonomy for any JSON value.} \label{fig:javascript-example}
\end{figure*}

The module exposes a single function that takes any JSON value and returns the
sequence of taxonomy qualifiers as an array of strings. Its usage is
exemplified in \autoref{fig:javascript-example}.

\subsection{CLI}

\begin{figure*}[ht!]
  \frame{\includegraphics[width=\linewidth]{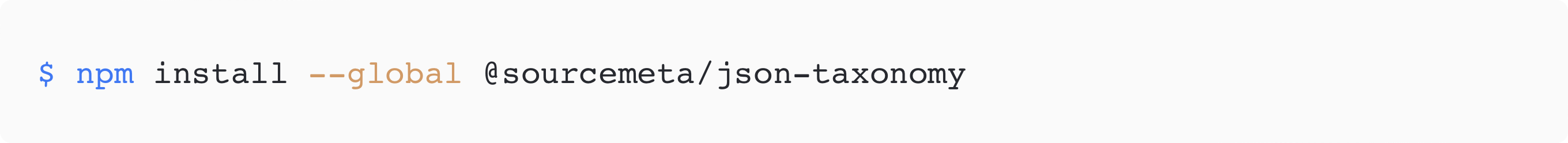}}
  \caption{The companion command-line interface program that implements the
  JSON Taxonomy is also installed through the npm package manager.}
\label{fig:npm-install-global} \end{figure*}

The npm package includes a simple command-line interface program that can be
globally installed as shown in \autoref{fig:npm-install-global}.

\begin{figure*}[ht!]
  \frame{\includegraphics[width=\linewidth]{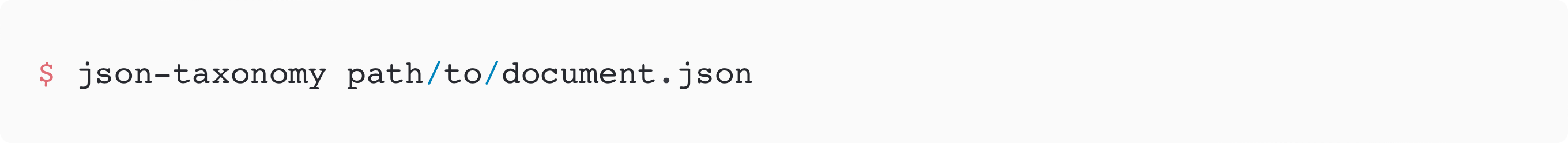}}
  \caption{An example of running the JSON Taxonomy command-line program with a
  fictitious JSON document file as input.} \label{fig:json-taxonomy-cli}
\end{figure*}

The CLI program takes the path to a JSON document as an argument and outputs
the taxonomy to standard output as shown in \autoref{fig:json-taxonomy-cli}.

\section{Demonstration}

In this section, we demonstrate various JSON documents for Tier 1, Tier 2 and
Tier 3 as per the taxonomy defined in \cite{viotti2022benchmark}. For each
demonstration, we provide the definition, characteristics and document
structure. These documents are as follows:

\begin{itemize}

  \item \textbf{Grunt.js Clean Task.} Tier 1 Minified $<$ 100 bytes, textual, redundant, flat. See \autoref{fig:example-gruntcleantask}.

  \item \textbf{CircleCI Matrix.} Tier 1 Minified $<$ 100 bytes, numeric, non-redundant, nested. See \autoref{fig:example-circlecimatrix}.

  \item \textbf{TSLint Linter (Basic).} Tier 1 Minified $<$ 100 bytes, boolean, non-redundant, nested. See \autoref{fig:example-tslintbasic}.

  \item \textbf{Entry Point Regulation Manifest.} Tier 2 Minified $\geq$ 100 $<$ 1000 bytes, textual, redundant, nested. See \autoref{fig:example-epr}.

  \item \textbf{TravisCI Notifications Configuration.} Tier 2 Minified $\geq$ 100 $<$ 1000 bytes, textual, redundant, flat. See \autoref{fig:example-travisnotifications}.

  \item \textbf{GeoJSON Example Document.} Tier 2 Minified $\geq$ 100 $<$ 1000 bytes, numeric, redundant, nested. See \autoref{fig:example-geojson}.

  \item \textbf{GitHub FUNDING Sponsorship (Empty).} Tier 2 Minified $\geq$ 100 $<$ 1000 bytes, boolean, redundant, flat. See \autoref{fig:example-github-funding}.

  \item \textbf{NPM Package.json Example Manifest.} Tier 3 Minified $\geq$ 1000 bytes, textual, non-redundant, flat. See \autoref{fig:example-package-json}.

  \item \textbf{JSON Resume Example.} Tier 3 Minified $\geq$ 1000 bytes, textual, non-redundant, nested. See \autoref{fig:example-jsonresume}.

  \item \textbf{ESLint Configuration Document.} Tier 3 Minified $\geq$ 1000 bytes, numeric, redundant, flat. See \autoref{fig:example-eslintrc}.

  \item \textbf{Nightwatch.js Test Framework Configuration.} Tier 3 Minified $\geq$ 1000 bytes, boolean, redundant, flat. See \autoref{fig:example-nightwatch}.

\end{itemize}

\clearpage
\subsection{Grunt.js Clean Task}

\begin{figure*}[ht!]
\frame{\includegraphics[width=\linewidth]{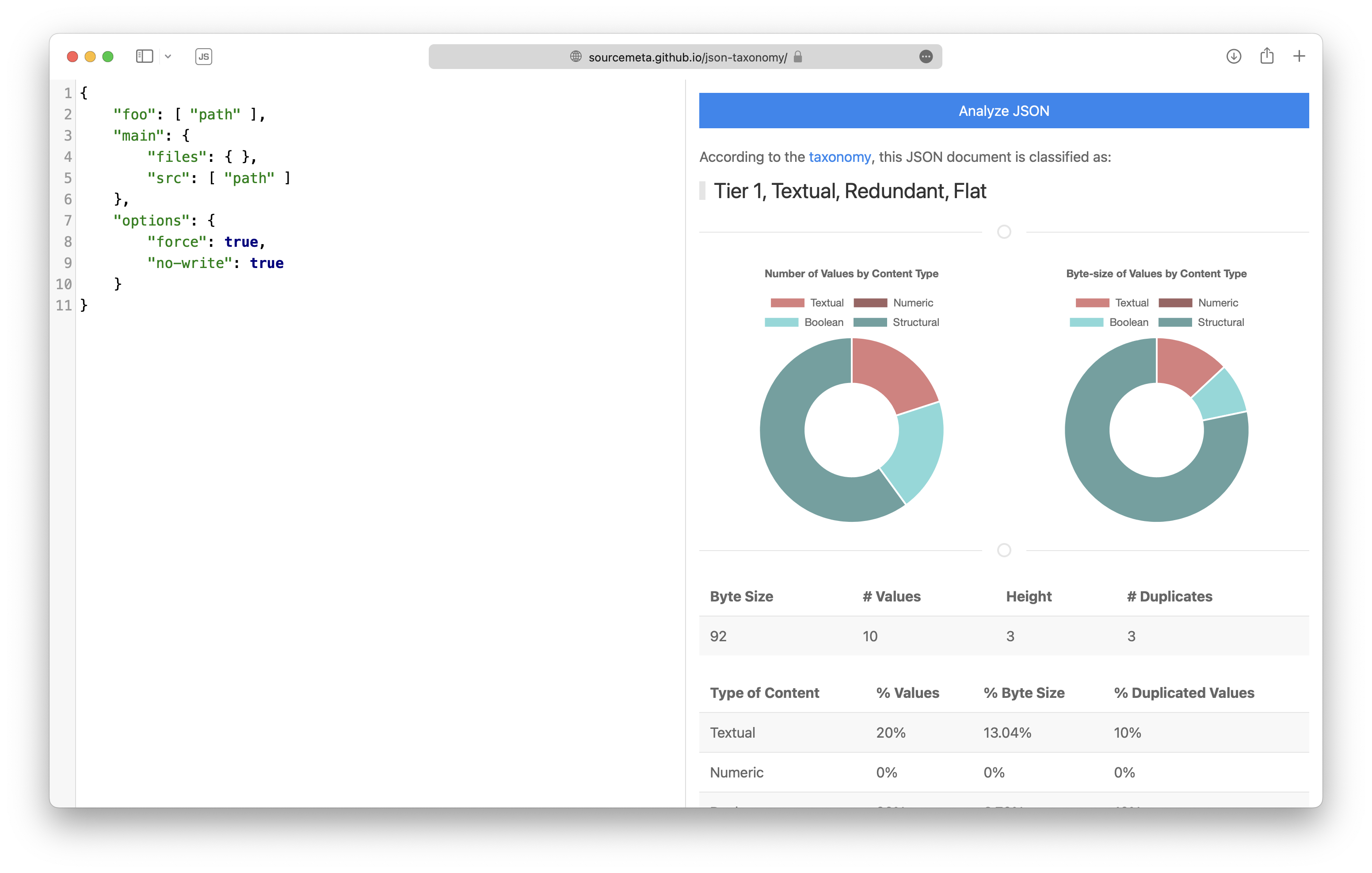}}
\caption{Grunt.js Clean Task} \label{fig:example-gruntcleantask}
\end{figure*}

\textbf{Definition.} Grunt.js \footnote{\url{https://gruntjs.com}} is an
open-source task runner for the JavaScript \cite{ECMA-262} programming language
used by a wide range of companies in the software development industry such as
Twitter, Adobe, and Mozilla
\footnote{\url{https://gruntjs.com/who-uses-grunt}}. In
\autoref{fig:example-gruntcleantask}, we demonstrate a \textbf{Tier 1 minified
$<$ 100 bytes textual redundant flat} (Tier 1 TRF from
\autoref{table:naming-conventions}) JSON document that consists of an example
configuration for a built-in plugin to clear files and folders called
\texttt{grunt-contrib-clean}
\footnote{\url{https://github.com/gruntjs/grunt-contrib-clean}}.

\textbf{Characteristics.} With a size of 92 bytes in UTF-8
\cite{UnicodeStandard} minified form, 10 values, a height
\cite{viotti2022benchmark} of 3 and 3 duplicated values, Grunt.js Contrib Clean
is a JSON document whose content is determined by structure both in terms of
number of values and cumulative byte-size.  It has 4 non-structural values: 2
textual values and 2 boolean values. However, the textual values are 4.3\%
larger in byte-size than the boolean values.

\textbf{Document Structure.} This document contains equal number of boolean and
textual values, but the cumulative byte-size of the textual values is larger
than the cumulative byte-size of the boolean values. Therefore, this document
is \emph{textual} according to the taxonomy \cite{viotti2022benchmark}.  The
boolean and string values in the document have a 50\% duplication.  Therefore,
this document is \emph{redundant} according to the taxonomy
\cite{viotti2022benchmark}.  The height of this document is 3 and its largest
level is 2.  Therefore, this document is \emph{flat} according to the taxonomy
\cite{viotti2022benchmark}.

\clearpage
\subsection{CircleCI Matrix}

\begin{figure*}[ht!]
\frame{\includegraphics[width=\linewidth]{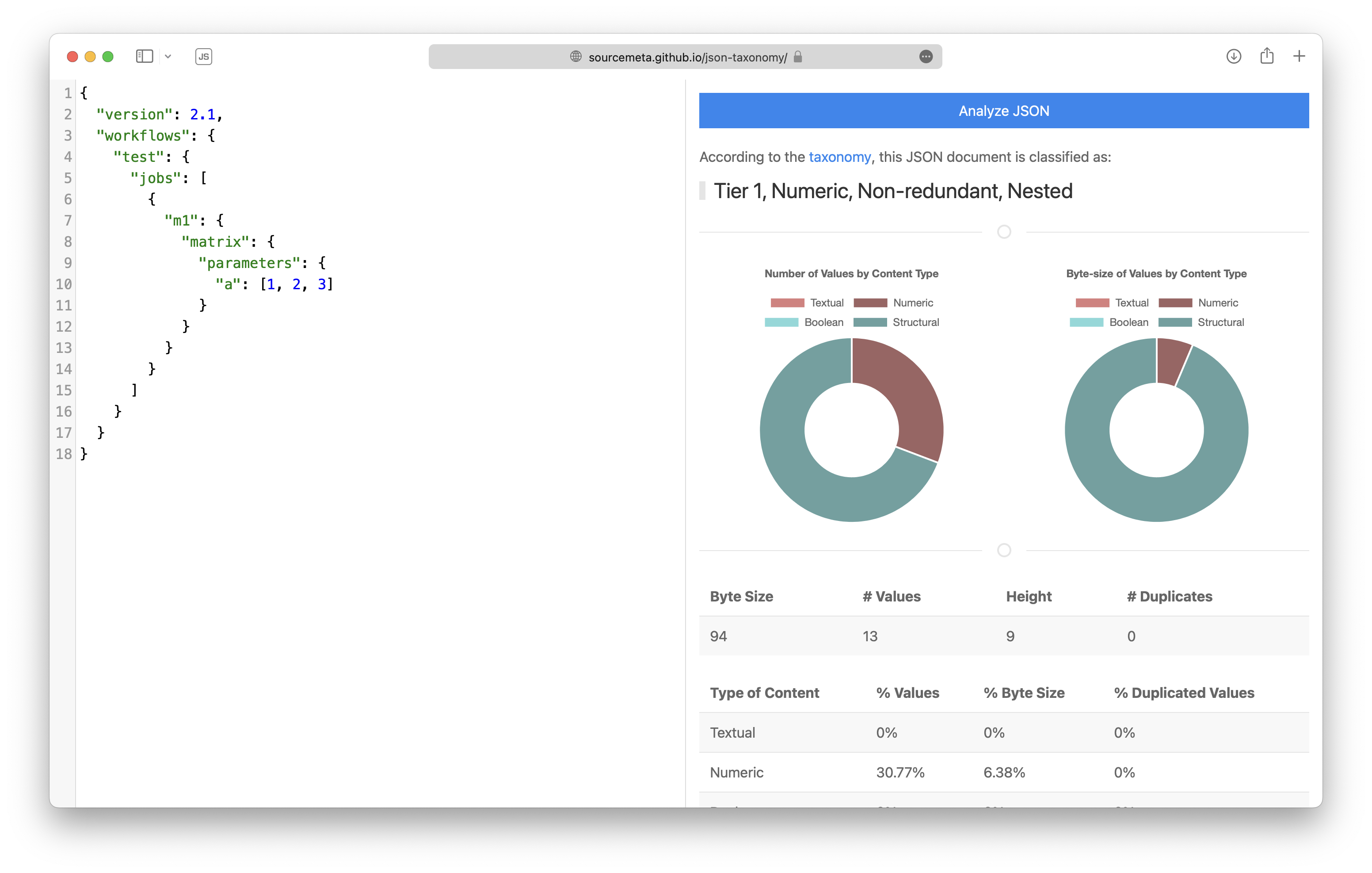}}
\caption{CircleCI Matrix} \label{fig:example-circlecimatrix}
\end{figure*}

\textbf{Definition.} CircleCI \footnote{\url{https://circleci.com}} is a
commercial cloud-provider of continuous integration and deployment pipelines
used by a wide range of companies in the software development industry such as
Facebook, Spotify, and Heroku \footnote{\url{https://circleci.com/customers/}}.
In \autoref{fig:example-circlecimatrix}, we demonstrate a \textbf{Tier 1
minified $<$ 100 bytes numeric non-redundant nested} (Tier 1 NNN from
\autoref{table:naming-conventions}) JSON document that represents a pipeline
configuration file for CircleCI that declares the desired CircleCI version and
defines a workflow that contains a single blank matrix-based job.

\textbf{Characteristics.} With a size of 94 bytes in UTF-8
\cite{UnicodeStandard} minified form, 13 values, a height
\cite{viotti2022benchmark} of 9 and no duplicated values, CircleCI Matrix is a
JSON document whose content is determined by structural values both in terms of
number and cumulative byte-size. It only has non-structural values that are
numeric and add up to less than 7\% of the total byte-size.

\textbf{Document Structure.} Leaving structural values aside, this document
only contains numeric values.  Therefore, this document is \emph{numeric}
according to the taxonomy \cite{viotti2022benchmark}.  None of the values in
this document are duplicated. Therefore, this document is \emph{non-redundant}
according to the taxonomy \cite{viotti2022benchmark}.  The height of this
document is 9 and its largest level is 9.  Therefore, this document is
\emph{nested} according to the taxonomy \cite{viotti2022benchmark}.

\clearpage
\subsection{TSLint Linter (Basic)}

\begin{figure*}[ht!]
\frame{\includegraphics[width=\linewidth]{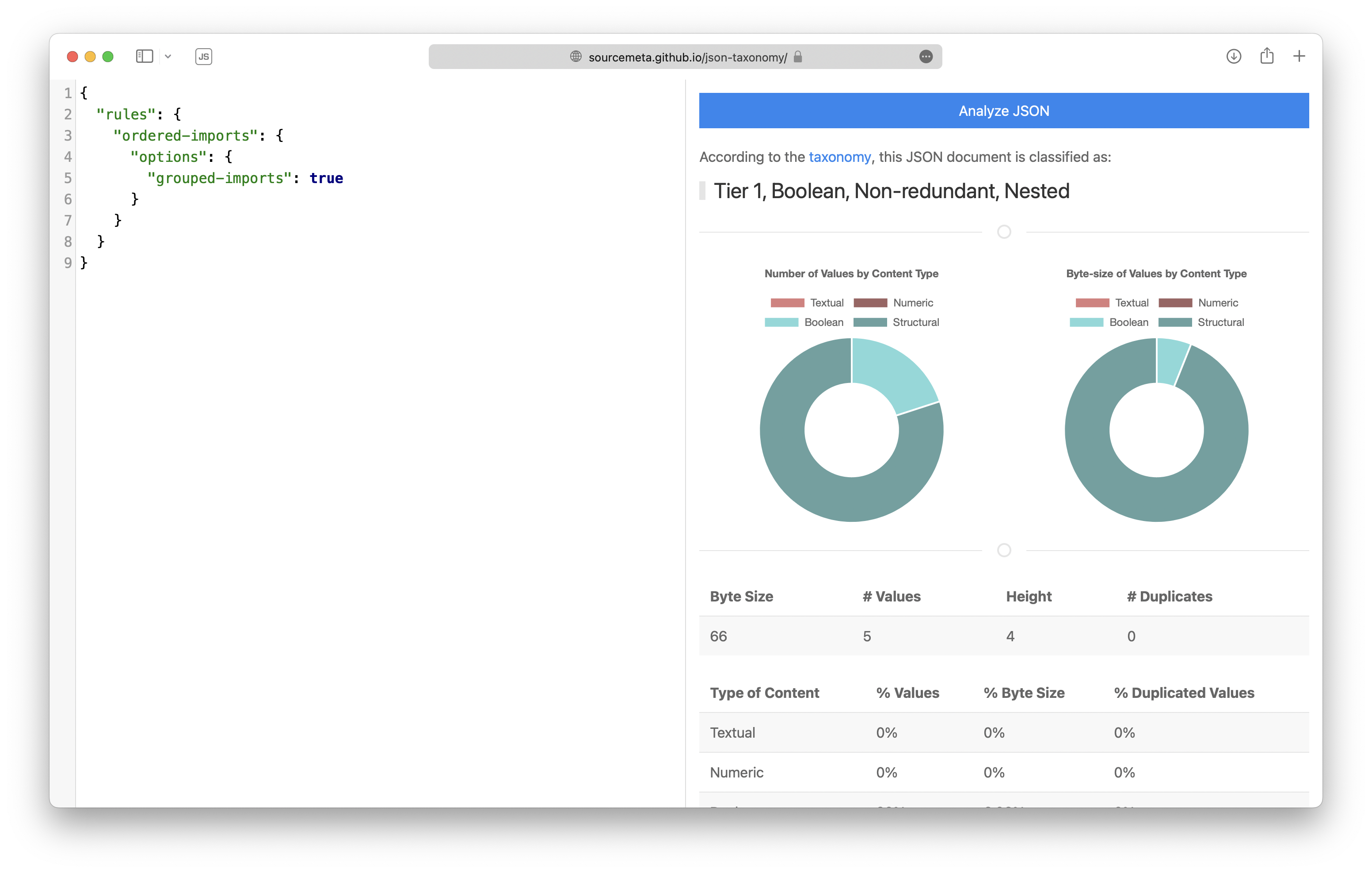}}
\caption{TSLint Linter (Basic)} \label{fig:example-tslintbasic}
\end{figure*}

\textbf{Definition.} TSLint \footnote{\url{https://palantir.github.io/tslint}}
is now an obsolete open-source linter for the TypeScript
\footnote{\url{https://www.typescriptlang.org}} programming language. TSLint
was created by the Big Data analytics company Palantir
\footnote{\url{https://www.palantir.com}} and was merged with the ESLint
open-source JavaScript linter in 2019
\footnote{\url{https://github.com/palantir/tslint/issues/4534}}. In
\autoref{fig:example-tslintbasic}, we demonstrate a \textbf{Tier 1 minified $<$
100 bytes boolean non-redundant nested} (Tier 1 BNN from
\autoref{table:naming-conventions}) JSON document that consists of a basic
TSLint configuration that enforces grouped alphabetized imports.

\textbf{Characteristics.} With a size of 66 bytes in UTF-8
\cite{UnicodeStandard} minified form, 5 values, a height
\cite{viotti2022benchmark} of 4 and no duplicated values, TSLint Linter (Basic)
is a JSON document whose content is determined by structural values both in
terms of number and cumulative byte-size. It has 1 non-structural value: a
boolean that add up to less than 7\% of the total byte-size.

\textbf{Document Structure.} Leaving structural values aside, this document
only contains boolean values.  Therefore, this document is \emph{boolean}
according to the taxonomy \cite{viotti2022benchmark}.  None of the values in
this document are duplicated. Therefore, this document is \emph{non-redundant}
according to the taxonomy \cite{viotti2022benchmark}.  The height of this
document is 4 and its largest level is 4.  Therefore, this document is
\emph{nested} according to the taxonomy \cite{viotti2022benchmark}.

\clearpage
\subsection{Entry Point Regulation Manifest}

\begin{figure*}[ht!]
\frame{\includegraphics[width=\linewidth]{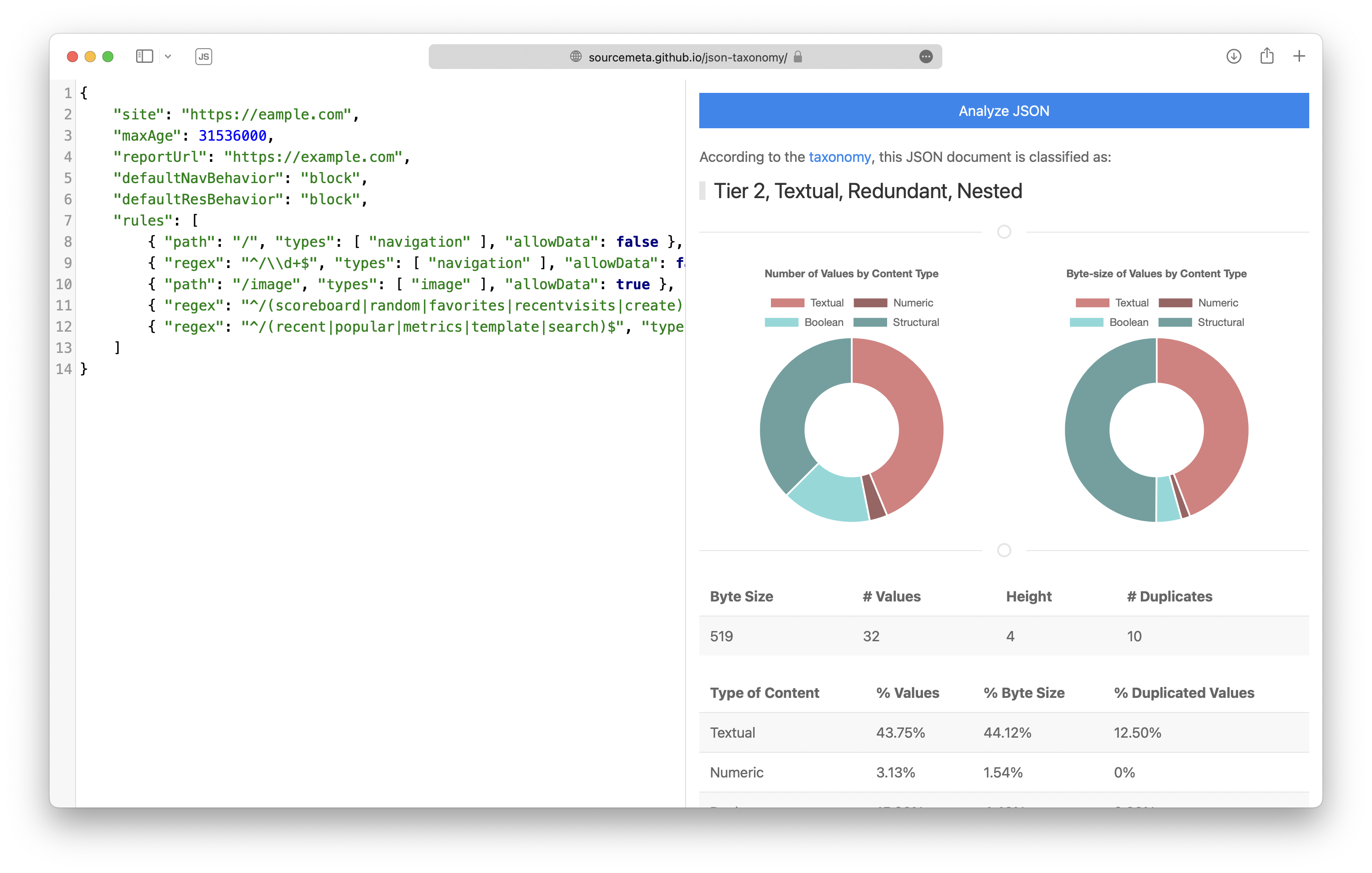}}
\caption{Entry Point Regulation Manifest} \label{fig:example-epr}
\end{figure*}

\textbf{Definition.} Entry Point Regulation (EPR) \cite{EPR} is a W3C proposal
led by Google that defines a manifest that protects websites against cross-site
scripting attacks by allowing the developer to mark the areas of the
application that can be externally referenced. EPR manifests are used in the
web industry. In \autoref{fig:example-epr}, we demonstrate a \textbf{Tier 2
minified $\geq$ 100 $<$ 1000 bytes textual redundant nested} (Tier 2 TRN from
\autoref{table:naming-conventions}) JSON document that defines an example EPR
policy for a fictitious website.

\textbf{Characteristics.} With a size of 519 bytes in UTF-8
\cite{UnicodeStandard} minified form, 32 values, a height
\cite{viotti2022benchmark} of 4 and 10 duplicated values, Entry Point
Regulation Manifest is a JSON document whose content is determined by textual
values in terms of number but structural values in terms of cumulative
byte-size. It has 20 non-structural values: 14 textual values, 1 numeric value
and 5 boolean values. Together, the 20 non-structural values represent 50\% of
the cumulative byte-size.

\textbf{Document Structure.} This document includes textual, numeric and
boolean values.  Leaving structural values aside, this document is dominated by
textual values in terms of number and cumulative byte-size.  Therefore, this
document is \emph{textual} according to the taxonomy
\cite{viotti2022benchmark}.  In this document, 3 out of 5 boolean values, 4 out
of 14 textual values and 3 out of 12 structural values are duplicates.
Therefore, 31.25\% of its values are duplicates, making this document
\emph{redundant} according to the taxonomy \cite{viotti2022benchmark}.  The
height of this document is 4 and its largest level is 3.  Therefore, this
document is \emph{nested} according to the taxonomy \cite{viotti2022benchmark}.

\clearpage
\subsection{TravisCI Notifications Configuration}

\begin{figure*}[ht!]
\frame{\includegraphics[width=\linewidth]{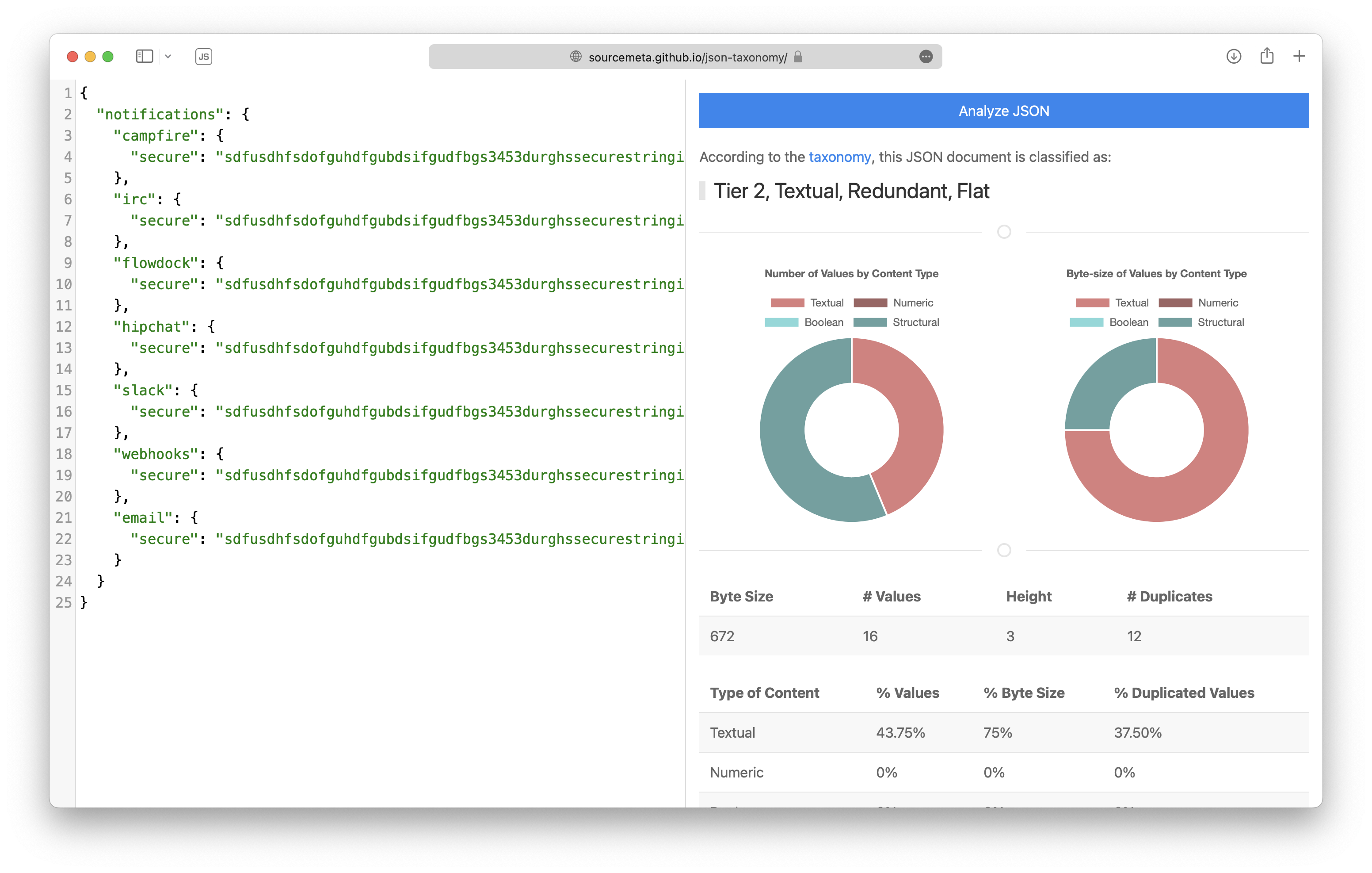}}
\caption{TravisCI Notifications Configuration} \label{fig:example-travisnotifications}
\end{figure*}

\textbf{Definition.} TravisCI \footnote{\url{https://travis-ci.com}} is a
commercial cloud-provider of continuous integration and deployment pipelines
used by a wide range of companies in the software development industry such as
ZenDesk, BitTorrent, and Engine Yard. In
\autoref{fig:example-travisnotifications}, we demonstrate a \textbf{Tier 2
minified $\geq$ 100 $<$ 1000 bytes textual redundant flat} (Tier 2 TRF from
\autoref{table:naming-conventions}) JSON document that consists of an example
pipeline configuration for TravisCI that declares a set of credentials to post
build notifications to various external services.

\textbf{Characteristics.} With a size of 672 bytes in UTF-8
\cite{UnicodeStandard} minified form, 16 values, a height
\cite{viotti2022benchmark} of 3 and 12 duplicated values, TravisCI
Notifications Configuration is a JSON document whose content is determined by
structural values in terms of number and textual values in terms of cumulative
byte-size. All of its 7 non-structural values are strings that add to 75\% of
the total byte-size.

\textbf{Document Structure.} Leaving structural values aside, this document
only contains textual values.  Therefore, this document is \emph{textual}
according to the taxonomy \cite{viotti2022benchmark}.  In this document, 6 out
of 7 textual values and 6 out of 9 structural values are duplicates. Therefore,
75\% of its values are duplicates, making this document \emph{redundant}
according to the taxonomy \cite{viotti2022benchmark}.  The height of this
document is 3 and its largest level is 3.  Therefore, this document is
\emph{flat} according to the taxonomy \cite{viotti2022benchmark}.

\clearpage
\subsection{GeoJSON Example Document}

\begin{figure*}[ht!]
\frame{\includegraphics[width=\linewidth]{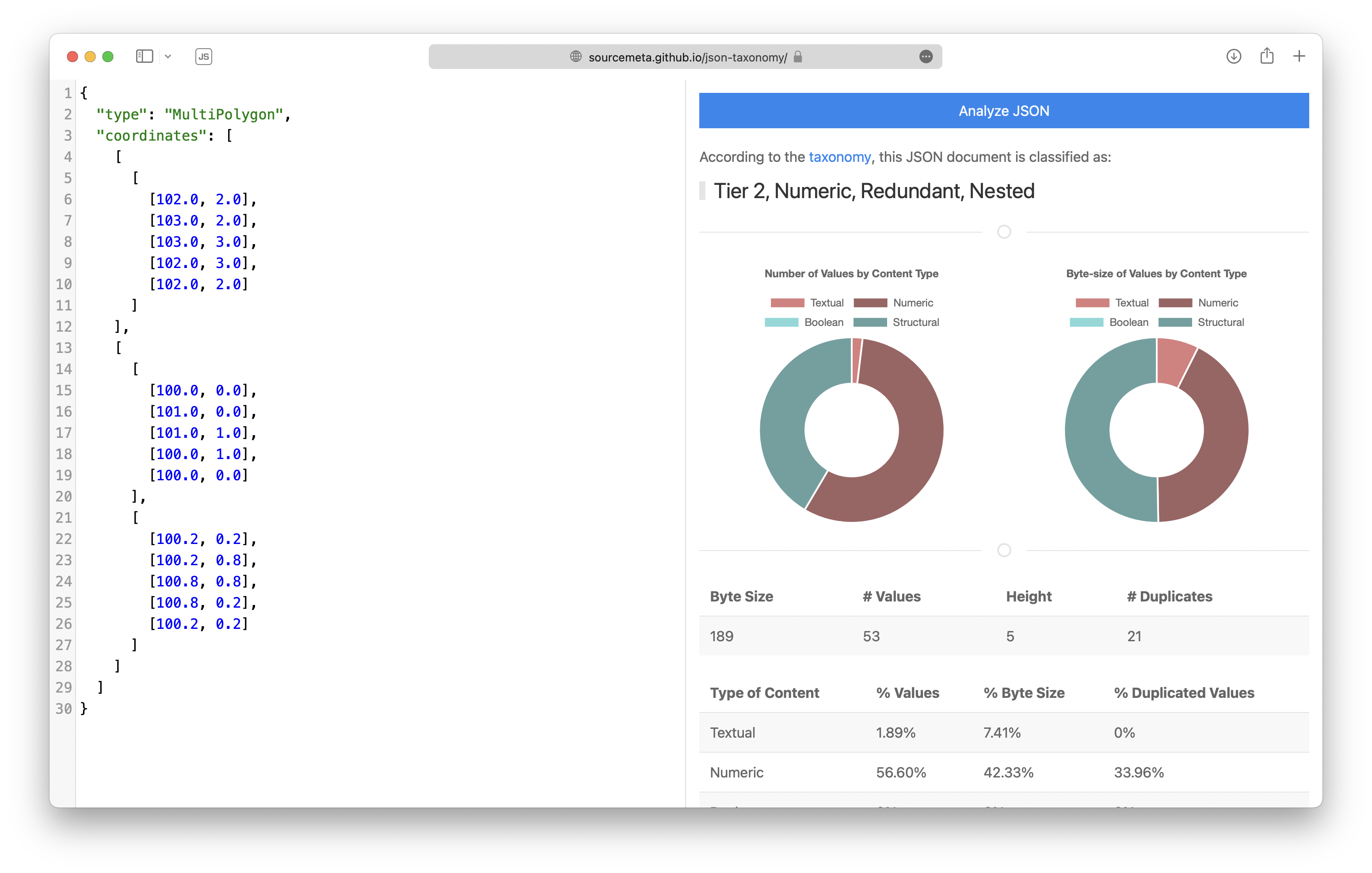}}
\caption{GeoJSON Example Document} \label{fig:example-geojson}
\end{figure*}

\textbf{Definition.} GeoJSON \cite{RFC7946} is a standard to encode geospatial
information using JSON. GeoJSON is used in industries that have geographical
and geospatial use cases such as engineering, logistics and telecommunications.
In \autoref{fig:example-geojson}, we demonstrate a \textbf{Tier 2 minified
$\geq$ 100 $<$ 1000 bytes numeric redundant nested} (Tier 2 NRN from
\autoref{table:naming-conventions}) JSON document that defines an example
polygon using the GeoJSON format.

\textbf{Characteristics.} With a size of 189 bytes in UTF-8
\cite{UnicodeStandard} minified form, 53 values, a height
\cite{viotti2022benchmark} of 5 and 21 duplicated values, GeoJSON Example
Document is a JSON document whose content is determined by numeric values in
terms of number but structural values in terms of cumulative byte-size. It has
31 non-structural values: 1 textual value and 30 numeric values. Together, the
31 non-structural values represent 49.7\% of the cumulative byte-size.

\textbf{Document Structure.} This document includes numeric and textual values.
Leaving structural values aside, this document is dominated by numeric values
in terms of number and cumulative byte-size.  Therefore, this document is
\emph{numeric} according to the taxonomy \cite{viotti2022benchmark}.  In this
document, 18 out of 30 numeric values and 3 out of 22 structural values are
duplicates. Therefore, 39.62\% of its values are duplicates, making this
document \emph{redundant} according to the taxonomy \cite{viotti2022benchmark}.
The height of this document is 5 and its largest level is 5.  Therefore, this
document is \emph{nested} according to the taxonomy \cite{viotti2022benchmark}.

\clearpage
\subsection{GitHub FUNDING Sponsorship (Empty)}

\begin{figure*}[ht!]
\frame{\includegraphics[width=\linewidth]{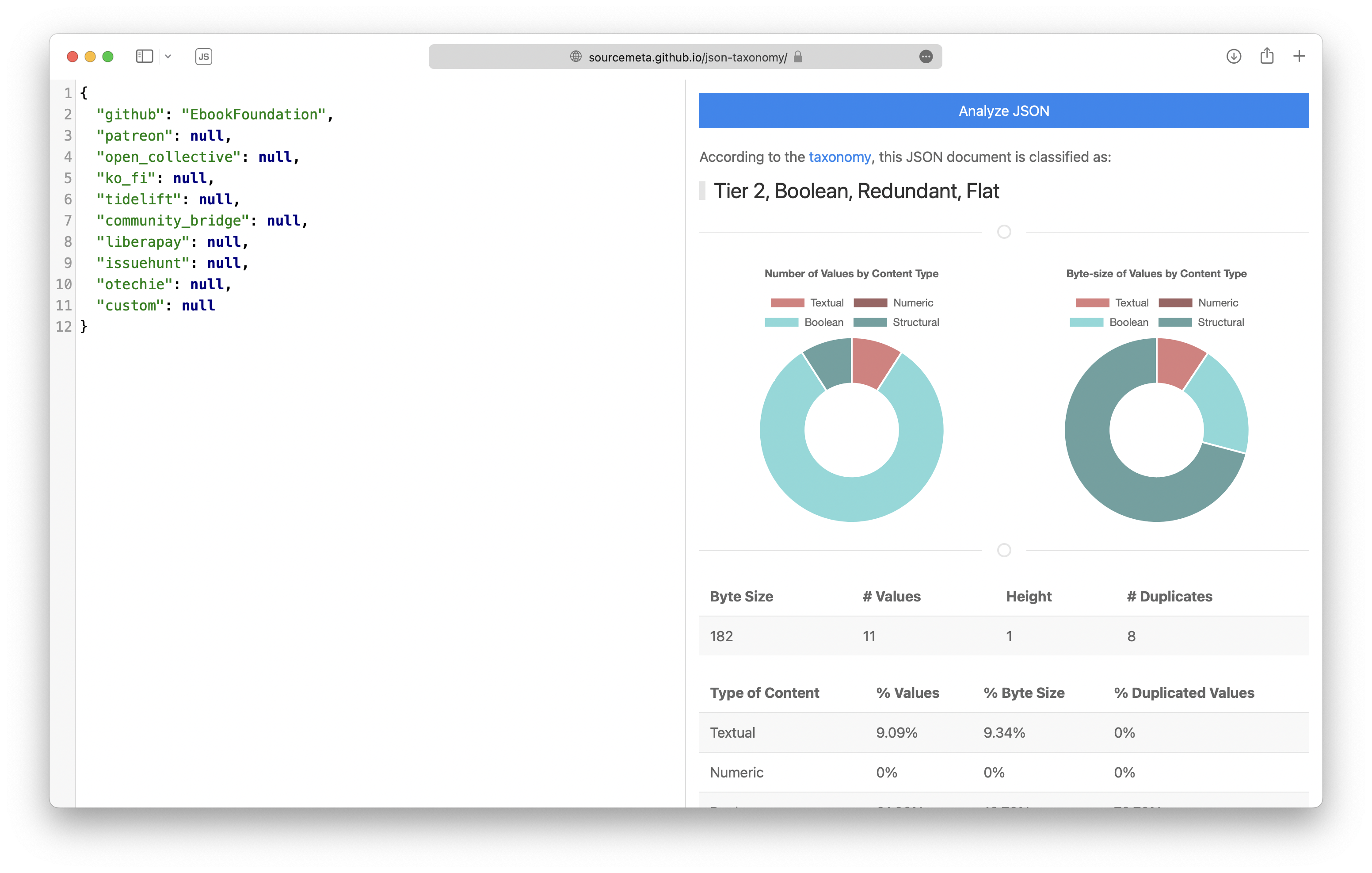}}
\caption{GitHub FUNDING Sponsorship (Empty)} \label{fig:example-github-funding}
\end{figure*}

\textbf{Definition.} The GitHub \footnote{\url{https://github.com}} software
hosting provider defines a \texttt{FUNDING}
\footnote{\url{https://docs.github.com/en/github/administering-a-repository/managing-repository-settings/displaying-a-sponsor-button-in-your-repository}}
file format to declare the funding platforms that an open-source project
supports. The \texttt{FUNDING} file format is used by the open-source software
industry. In \autoref{fig:example-github-funding}, we demonstrate a
\textbf{Tier 2 minified $\geq$ 100 $<$ 1000 bytes boolean redundant flat} (Tier
2 BRF from \autoref{table:naming-conventions}) JSON document that consists of a
definition that does not declare any supported funding platforms.

\textbf{Characteristics.} With a size of 182 bytes in UTF-8
\cite{UnicodeStandard} minified form, 11 values, a height
\cite{viotti2022benchmark} of 1 and 8 duplicated values, GitHub FUNDING
Sponsorship (Empty) is a JSON document whose content is determined by boolean
values in terms of number but structural values in terms of cumulative
byte-size. It has 10 non-structural values: 1 textual value and 9 boolean
values. Together, the 10 non-structural values represent 29.1\% of the
cumulative byte-size.

\textbf{Document Structure.} This document is dominated by boolean values in
terms of number.  Leaving structural values aside, this document is dominated
by boolean values in terms of cumulative byte-size.  Therefore, this document
is \emph{boolean} according to the taxonomy \cite{viotti2022benchmark}.  In
this document, 8 out of 9 boolean values are duplicates. Therefore, 72.73\% of
its values are duplicates, making this document \emph{redundant} according to
the taxonomy \cite{viotti2022benchmark}.  The height of this document is 1 and
its largest level is 1.  Therefore, this document is \emph{flat} according to
the taxonomy \cite{viotti2022benchmark}.

\clearpage
\subsection{NPM Package.json Example Manifest}

\begin{figure*}[ht!]
\frame{\includegraphics[width=\linewidth]{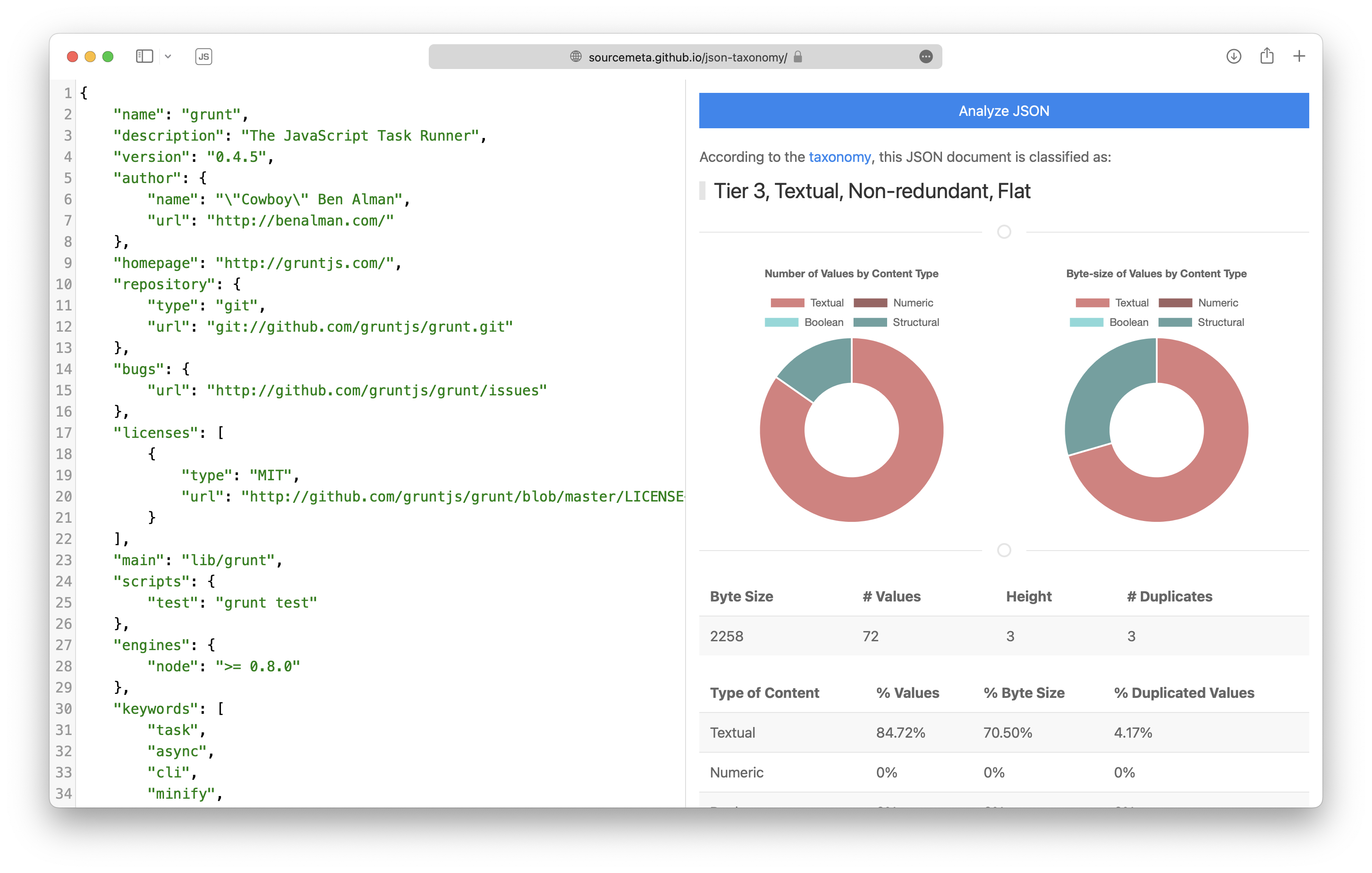}}
\caption{NPM Package.json Example Manifest} \label{fig:example-package-json}
\end{figure*}

\textbf{Definition.} Node.js Package Manager (NPM)
\footnote{\url{https://www.npmjs.com}} is an open-source package manager for
Node.js \footnote{\url{https://nodejs.org}}, a JavaScript \cite{ECMA-262}
runtime targetted at the web development industry. A package that is published
to NPM is declared using a JSON file called \texttt{package.json}
\footnote{\url{https://docs.npmjs.com/cli/v6/configuring-npm/package-json}}.
In \autoref{fig:example-package-json}, we demonstrate a \textbf{Tier 3 minified
$\geq$ 1000 bytes textual non-redundant flat} (Tier 3 TNF from
\autoref{table:naming-conventions}) JSON document that consists of a
\texttt{package.json} manifest that declares a particular version of the
Grunt.js \footnote{\url{https://gruntjs.com}} task runner.

\textbf{Characteristics.} With a size of 2258 bytes in UTF-8
\cite{UnicodeStandard} minified form, 72 values, a height
\cite{viotti2022benchmark} of 3 and 3 duplicated values, NPM Package.json
Example Manifest is a JSON document whose content is determined by textual
values in terms of both number and cumulative byte-size. It has 61
non-structural values. Together, the non-structural values represent 84.72\% of
the cumulative byte-size.

\textbf{Document Structure.} Leaving structural values aside, this document
only contains textual values.  Therefore, this document is \emph{textual}
according to the taxonomy \cite{viotti2022benchmark}.  In this document, 3 out
of 61 textual values are duplicates. Therefore, only 4.17\% of its values are
duplicates, making this document \emph{non-redundant} according to the taxonomy
\cite{viotti2022benchmark}.  The height of this document is 3 and its largest
level is 1.  Therefore, this document is \emph{flat} according to the taxonomy
\cite{viotti2022benchmark}.

\clearpage
\subsection{JSON Resume Example}

\begin{figure*}[ht!]
\frame{\includegraphics[width=\linewidth]{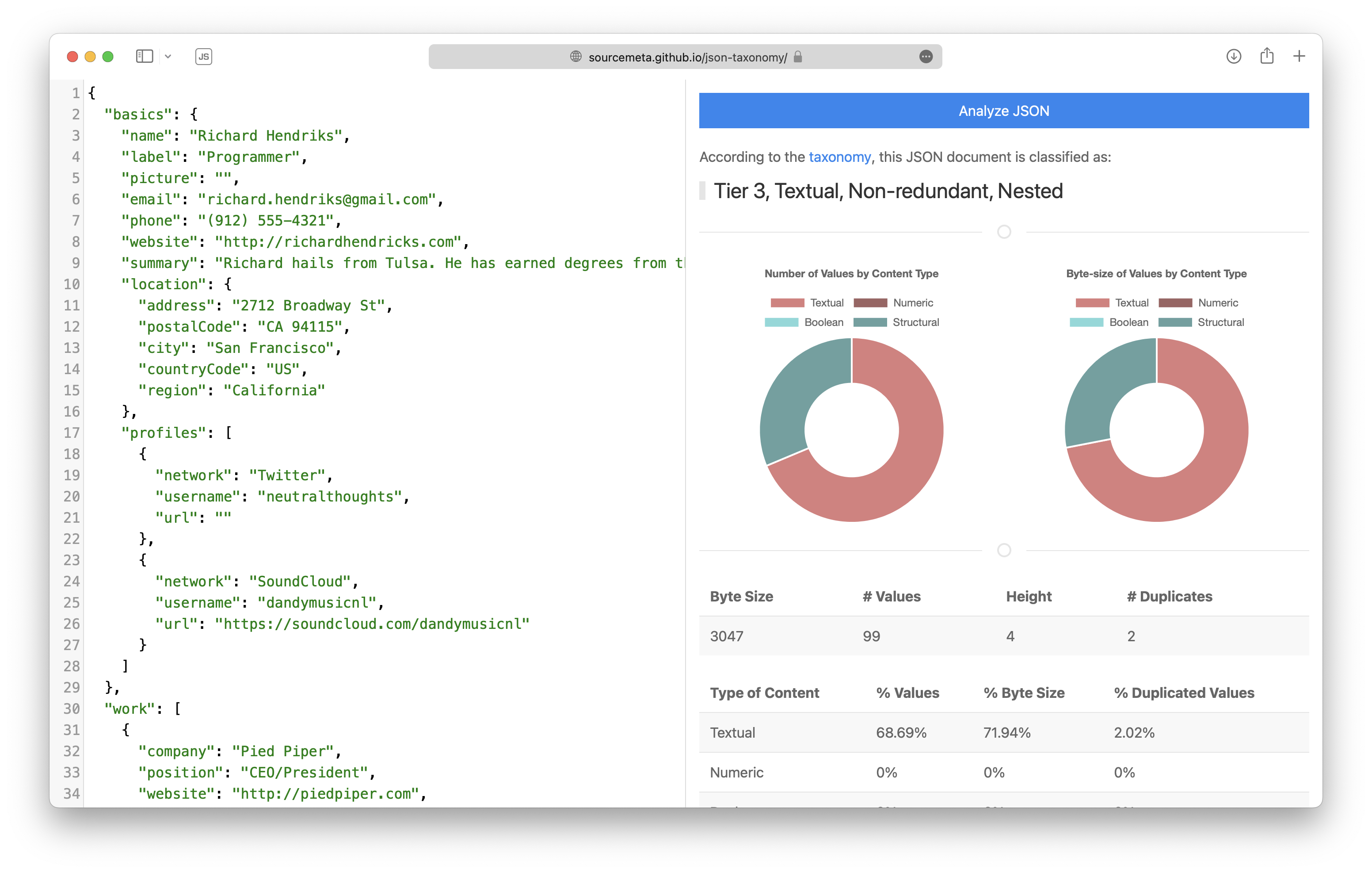}}
\caption{JSON Resume Example} \label{fig:example-jsonresume}
\end{figure*}

\textbf{Definition.} JSON Resume \footnote{\url{https://jsonresume.org}} is a
community-driven proposal for a JSON-based file format that declares and
renders themable resumes used in the recruitment industry. In
\autoref{fig:example-jsonresume}, we demonstrate a \textbf{Tier 3 minified
$\geq$ 1000 bytes textual non-redundant nested} (Tier 3 TNN from
\autoref{table:naming-conventions}) JSON document that consists of a detailed
example resume for a fictitious software programmer.

\textbf{Characteristics.} With a size of 3047 bytes in UTF-8
\cite{UnicodeStandard} minified form, 99 values, a height
\cite{viotti2022benchmark} of 4 and 2 duplicated values, JSON Resume Example is
a JSON document whose content is determined by textual values in terms of both
number and cumulative byte-size. It has 68 non-structural values. Together, the
non-structural values represent 68.69\% of the cumulative byte-size.

\textbf{Document Structure.} Leaving structural values aside, this document
only contains textual values.  Therefore, this document is \emph{textual}
according to the taxonomy \cite{viotti2022benchmark}.  In this document, 2 out
of 68 textual values are duplicates. Therefore, only 2.02\% of its values are
duplicates, making this document \emph{non-redundant} according to the taxonomy
\cite{viotti2022benchmark}.  The height of this document is 4 and its largest
level is 3.  Therefore, this document is \emph{nested} according to the
taxonomy \cite{viotti2022benchmark}.

\clearpage
\subsection{ESLint Configuration Document}

\begin{figure*}[ht!]
\frame{\includegraphics[width=\linewidth]{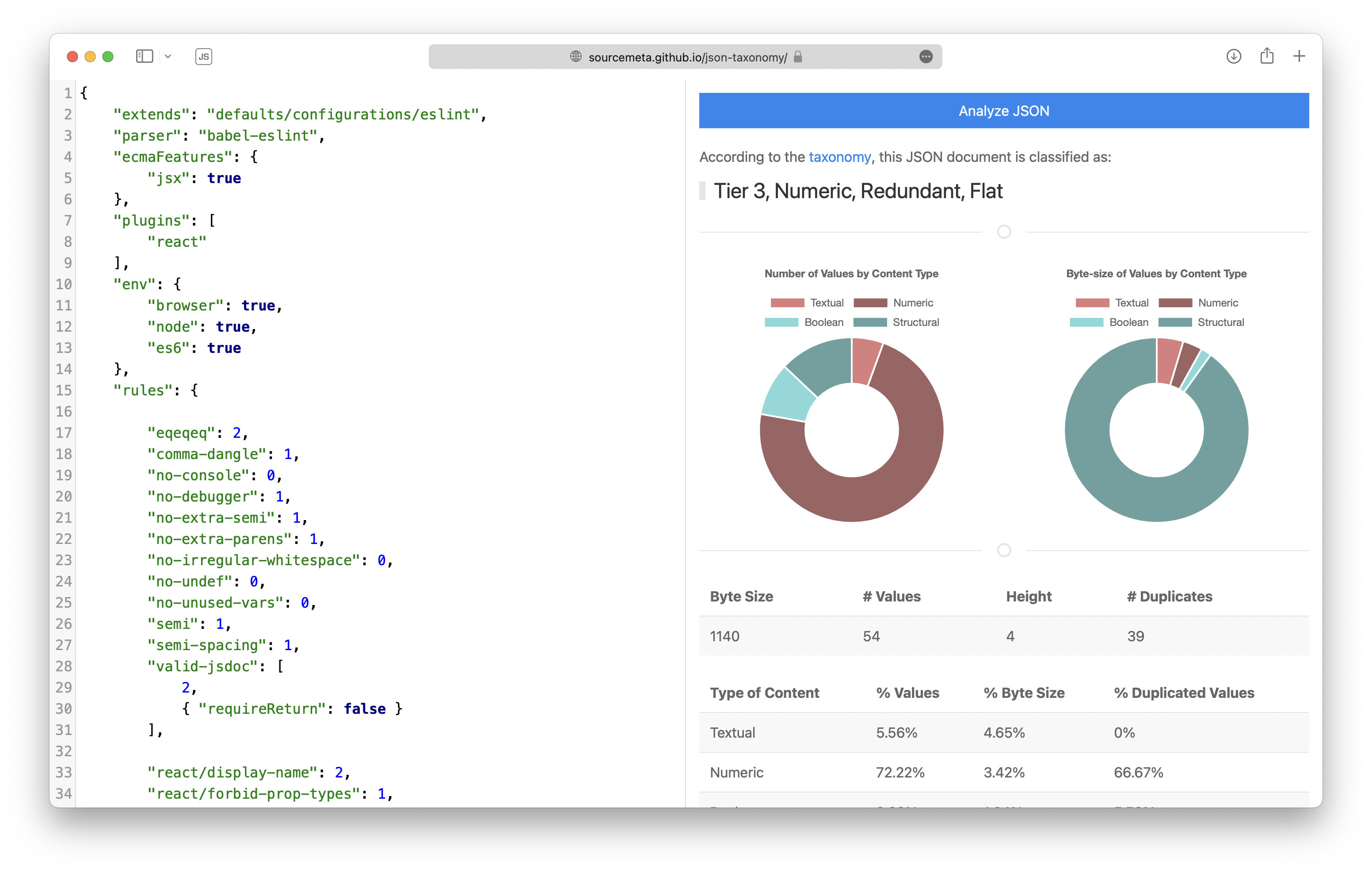}}
\caption{ESLint Configuration Document} \label{fig:example-eslintrc}
\end{figure*}

\textbf{Definition.} ESLint \footnote{\url{https://eslint.org}} is a popular
open-source extensible linter for the JavaScript \cite{ECMA-262} programming
language used by a wide range of companies in the software development industry
such as Google, Salesforce, and Airbnb. In \autoref{fig:example-eslintrc}, we
demonstrate a \textbf{Tier 3 minified $\geq$ 1000 bytes numeric redundant flat}
(Tier 3 NRF from \autoref{table:naming-conventions}) JSON document that defines
a browser and Node.js linter configuration that defines general-purposes and
\emph{React.js}-specific \footnote{\url{https://reactjs.org}} linting rules.

\textbf{Characteristics.} With a size of 1140 bytes in UTF-8
\cite{UnicodeStandard} minified form, 54 values, a height
\cite{viotti2022benchmark} of 4 and 39 duplicated values, ESLint Configuration
Document is a JSON document whose content is determined by numeric values in
terms of number but structural values in terms of cumulative byte-size. It has
47 non-structural values: 3 textual values, 39 numeric values and 5 boolean
values.  Together, the 47 non-structural values represent only 9.91\% of the
cumulative byte-size.

\textbf{Document Structure.} This document includes textual, numeric and
boolean values.  While this document is dominated by numeric values in terms of
number, numeric values do not make for the majority of cumulative byte-size by
content type.  However, the high difference in the number of numeric values and
other types of values increases the numeric weight of the document and makes
the document \emph{numeric} according to the taxonomy
\cite{viotti2022benchmark}.  In this document, 36 out of 39 numeric values and
3 out of 5 boolean values are duplicates. Therefore, 72.23\% of its values are
duplicates, making this document \emph{redundant} according to the taxonomy
\cite{viotti2022benchmark}.  The height of this document is 4 and its largest
level is 2.  Therefore, this document is \emph{flat} according to the taxonomy
\cite{viotti2022benchmark}.

\clearpage
\subsection{Nightwatch.js Test Framework Configuration}

\begin{figure*}[ht!]
\frame{\includegraphics[width=\linewidth]{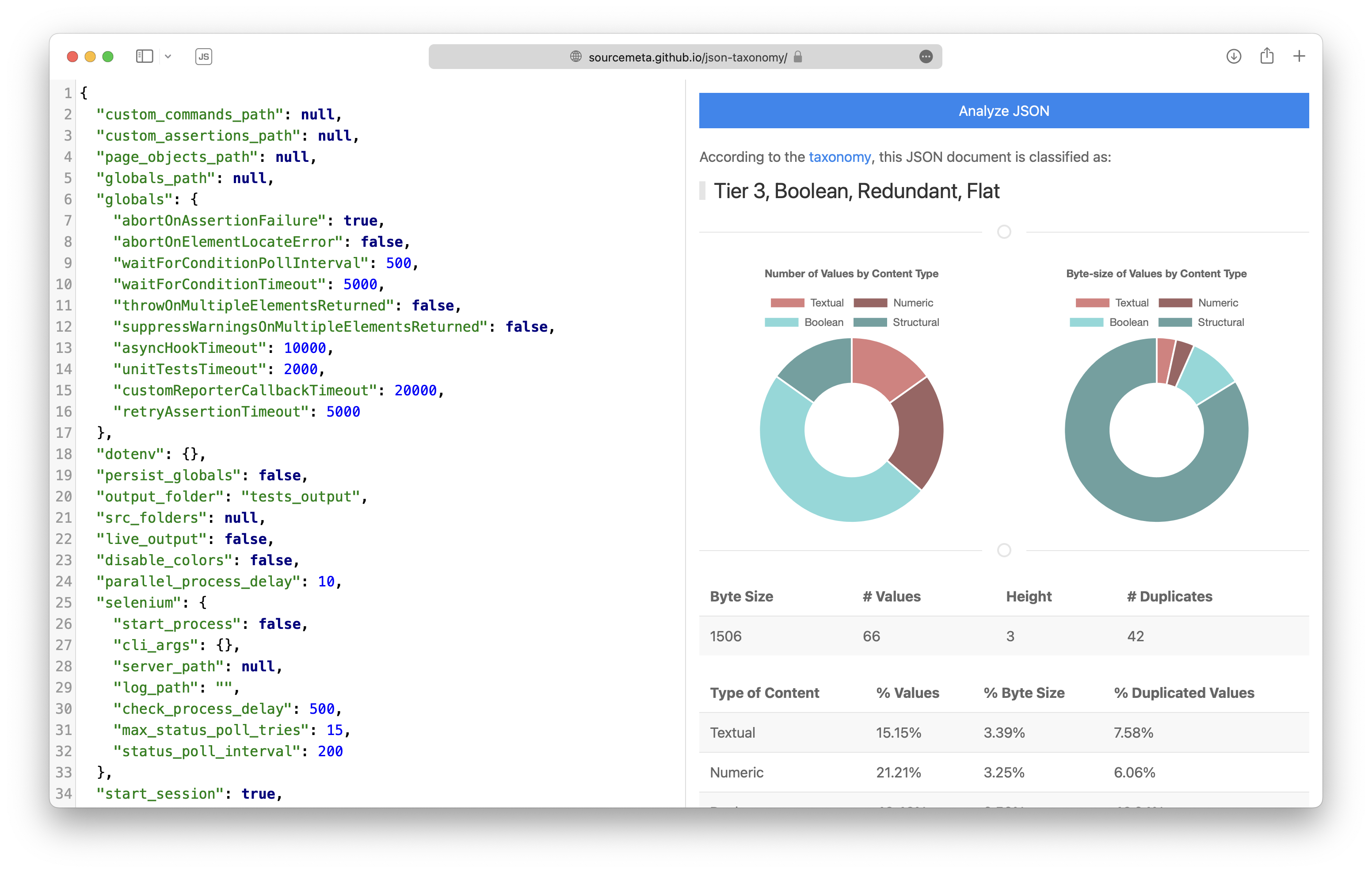}}
\caption{Nightwatch.js Test Framework Configuration} \label{fig:example-nightwatch}
\end{figure*}

\textbf{Definition.} Nightwatch.js \footnote{\url{https://nightwatchjs.org}} is
an open-source browser automation solution used in the software testing
industry. In \autoref{fig:example-nightwatch}, we demonstrate a \textbf{Tier 3
minified $\geq$ 1000 bytes boolean redundant flat} (Tier 3 BRF from
\autoref{table:naming-conventions}) JSON document that consists of a
Nightwatch.js configuration file that defines a set of general-purpose
WebDriver \cite{webdriver} and Selenium
\footnote{\url{https://www.selenium.dev}} options.

\textbf{Characteristics.} With a size of 1506 bytes in UTF-8
\cite{UnicodeStandard} minified form, 66 values, a height
\cite{viotti2022benchmark} of 3 and 42 duplicated values, Nightwatch.js Test
Framework Configuration is a JSON document whose content is determined by
boolean values in terms of number but structural values in terms of cumulative
byte-size. It has 56 non-structural values: 10 textual values, 14 numeric
values and 32 boolean values.  Together, the 56 non-structural values represent
only 16.2\% of the cumulative byte-size.

\textbf{Document Structure.} This document includes textual, numeric and
boolean values.  Leaving structural values aside, this document is dominated by
boolean values in terms of number and cumulative byte-size.  Therefore, this
document is \emph{boolean} according to the taxonomy
\cite{viotti2022benchmark}.  In this document, 5 out of 10 textual values, 4
out of 14 numeric values, 29 out of 32 boolean values and 4 out of 10
structural values are duplicates.  Therefore, 63.64\% of its values are
duplicates, making this document \emph{redundant} according to the taxonomy
\cite{viotti2022benchmark}.  The height of this document is 3 and its largest
level is 1.  Therefore, this document is \emph{flat} according to the taxonomy
\cite{viotti2022benchmark}.

\clearpage
\section{Conclusions and Further Work}

In this paper, we present JSON Stats Analyzer, a free-to-use open-source
web-based JavaScript tool that implements our proposed JSON taxonomy
\cite{viotti2022benchmark} that categorizes JSON documents according to their
size, content, redundancy and nesting characteristics.

There are several avenues to extend our work: a visualization dashboard that
compares multiple JSON documents across the tiers based on their content type,
redundancy characteristics and document structure providing us with a meta
analysis of how the documents with a variety of characteristics fare across the
tiers; building a wider dataset of real-world JSON documents that demonstrates
the characteristics pertinent to each tier. We welcome contributions to build
this dataset from the industry professionals and academic community alike.

\bibliographystyle{ACM-Reference-Format}
\bibliography{jsonstats}


\begin{thebibliography}{8}


\ifx \showCODEN    \undefined \def \showCODEN     #1{\unskip}     \fi
\ifx \showDOI      \undefined \def \showDOI       #1{#1}\fi
\ifx \showISBNx    \undefined \def \showISBNx     #1{\unskip}     \fi
\ifx \showISBNxiii \undefined \def \showISBNxiii  #1{\unskip}     \fi
\ifx \showISSN     \undefined \def \showISSN      #1{\unskip}     \fi
\ifx \showLCCN     \undefined \def \showLCCN      #1{\unskip}     \fi
\ifx \shownote     \undefined \def \shownote      #1{#1}          \fi
\ifx \showarticletitle \undefined \def \showarticletitle #1{#1}   \fi
\ifx \showURL      \undefined \def \showURL       {\relax}        \fi
\providecommand\bibfield[2]{#2}
\providecommand\bibinfo[2]{#2}
\providecommand\natexlab[1]{#1}
\providecommand\showeprint[2][]{arXiv:#2}

\bibitem[Butler et~al\mbox{.}(2016)]%
        {RFC7946}
\bibfield{author}{\bibinfo{person}{H. Butler}, \bibinfo{person}{M. Daly},
  \bibinfo{person}{A. Doyle}, \bibinfo{person}{S. Gillies}, \bibinfo{person}{S.
  Hagen}, {and} \bibinfo{person}{T. Schaub}.} \bibinfo{year}{2016}\natexlab{}.
\newblock \bibinfo{booktitle}{\emph{The GeoJSON Format}}.
\newblock \bibinfo{type}{RFC}. \bibinfo{institution}{IETF}.
\newblock
\urldef\tempurl%
\url{https://doi.org/10.17487/RFC7946}
\showDOI{\tempurl}


\bibitem[Consortium(2019)]%
        {UnicodeStandard}
\bibfield{author}{\bibinfo{person}{The~Unicode Consortium}.}
  \bibinfo{year}{2019}\natexlab{}.
\newblock \bibinfo{booktitle}{\emph{The Unicode Standard, Version 12.1.0}}.
\newblock \bibinfo{type}{Standard}. \bibinfo{institution}{The Unicode
  Consortium}, \bibinfo{address}{Mountain View, CA}.
\newblock


\bibitem[{ECMA}(2017)]%
        {ECMA-404}
\bibfield{author}{\bibinfo{person}{{ECMA}}.} \bibinfo{year}{2017}\natexlab{}.
\newblock \bibinfo{booktitle}{\emph{{ECMA-404}: The JSON Data Interchange
  Syntax}}.
\newblock \bibinfo{publisher}{ECMA}, \bibinfo{address}{Geneva, CH}.
\newblock
\urldef\tempurl%
\url{https://www.ecma-international.org/publications/standards/Ecma-404.htm}
\showURL{%
\tempurl}


\bibitem[{ECMA}(2021)]%
        {ECMA-262}
\bibfield{author}{\bibinfo{person}{{ECMA}}.} \bibinfo{year}{2021}\natexlab{}.
\newblock \bibinfo{booktitle}{\emph{{ECMA-262}: ECMAScript 2021 language
  specification}}.
\newblock \bibinfo{publisher}{ECMA}, \bibinfo{address}{Geneva, CH}.
\newblock
\urldef\tempurl%
\url{https://www.ecma-international.org/publications/standards/Ecma-262.htm}
\showURL{%
\tempurl}


\bibitem[Ross and West(2016)]%
        {EPR}
\bibfield{author}{\bibinfo{person}{David Ross} {and} \bibinfo{person}{Mike
  West}.} \bibinfo{year}{2016}\natexlab{}.
\newblock \bibinfo{booktitle}{\emph{Entry Point Regulation}}.
\newblock \bibinfo{type}{{W3C} Working Group Note}. \bibinfo{institution}{W3C}.
\newblock
\newblock
\shownote{https://w3c.github.io/webappsec-epr/}.


\bibitem[Stewart and Burns(2020)]%
        {webdriver}
\bibfield{author}{\bibinfo{person}{Simon Stewart} {and} \bibinfo{person}{David
  Burns}.} \bibinfo{year}{2020}\natexlab{}.
\newblock \bibinfo{booktitle}{\emph{WebDriver}}.
\newblock \bibinfo{type}{{W3C} Working Draft}. \bibinfo{institution}{W3C}.
\newblock


\bibitem[Vargas et~al\mbox{.}(2019)]%
        {10.1145/3355369.3355594}
\bibfield{author}{\bibinfo{person}{Santiago Vargas}, \bibinfo{person}{Utkarsh
  Goel}, \bibinfo{person}{Moritz Steiner}, {and} \bibinfo{person}{Aruna
  Balasubramanian}.} \bibinfo{year}{2019}\natexlab{}.
\newblock \showarticletitle{Characterizing JSON Traffic Patterns on a CDN}. In
  \bibinfo{booktitle}{\emph{Proceedings of the Internet Measurement
  Conference}} (Amsterdam, Netherlands) \emph{(\bibinfo{series}{IMC '19})}.
  \bibinfo{publisher}{Association for Computing Machinery},
  \bibinfo{address}{New York, NY, USA}, \bibinfo{pages}{195–201}.
\newblock
\showISBNx{9781450369480}
\urldef\tempurl%
\url{https://doi.org/10.1145/3355369.3355594}
\showDOI{\tempurl}


\bibitem[Viotti and Kinderkhedia(2022)]%
        {viotti2022benchmark}
\bibfield{author}{\bibinfo{person}{Juan~Cruz Viotti} {and}
  \bibinfo{person}{Mital Kinderkhedia}.} \bibinfo{year}{2022}\natexlab{}.
\newblock \bibinfo{title}{A Benchmark of JSON-compatible Binary Serialization
  Specifications}.
\newblock
\newblock
\showeprint[arxiv]{2201.03051}~[cs.SE]


\end{thebibliography}

\end{document}